\documentclass[letterpaper, 10 pt, conference]{ieeeconf}  

\IEEEoverridecommandlockouts                              
\usepackage{amsthm}

\usepackage{cite}

\usepackage{amsmath,amssymb,amsfonts}
\usepackage{algorithmic}
\usepackage[ruled,linesnumbered,noend,algo2e]{algorithm2e}
\usepackage{algorithm}
\usepackage{graphicx}
\usepackage{textcomp}

\usepackage[compact]{titlesec}        
\usepackage[svgnames]{xcolor}

\usepackage{tcolorbox}
\usepackage{colortbl,array}
 
\DeclareMathOperator*{\argmin}{arg\,min}

\newtheorem*{problem*}{Problem}
\definecolor{shadecolor}{named}{LightGray}
\newtheorem{theorem}{Theorem}
\newtheorem{remark}{Remark}
\newtheorem{assumption}[theorem]{Assumption}

\usepackage{url}
\makeatletter
\newcommand*{\rom}[1]{\expandafter\@slowromancap\romannumeral #1@}
\makeatother
\def\BibTeX{{\rm B\kern-.05em{\sc i\kern-.025em b}\kern-.08em
    T\kern-.1667em\lower.7ex\hbox{E}\kern-.125emX}}

\newcommand{\RNum}[1]{\uppercase\expandafter{\romannumeral #1\relax}}

\title{\LARGE \bf
On the Interplay of Privacy, Persuasion and Quantization
}

\author{Anju Anand and Emrah Akyol
\thanks{Authors are with the Binghamton University--SUNY, Binghamton, NY,
13902 USA {\tt\small {\{aanand6,eakyol\}}@binghamton.edu.} This research is supported by the NSF via grants CCF \#1910715 and CAREER \#2048042.}
}

\begin{document}

\maketitle
\thispagestyle{empty}
\pagestyle{empty}

\begin{abstract}
 We develop a communication-theoretic framework for privacy-aware and resilient decision making in cyber-physical systems under \emph{misaligned} objectives between the encoder and the decoder. The encoder observes two correlated signals $(X,\theta)$ and transmits a finite-rate message $Z$ to aid a legitimate controller (the decoder) in estimating $X+\theta,$ while an eavesdropper intercepts $Z$ to infer the private parameter $\theta.$ Unlike conventional setups where encoder and decoder share a common MSE objective, here the encoder minimizes a Lagrangian that balances legitimate control fidelity \emph{and} the privacy leakage about $\theta.$ In contrast, the decoder’s goal is purely to minimize its own estimation error without regard for privacy. We analyze fully, partially, and non-revealing strategies that arise from this conflict, and characterize optimal linear encoders when the rate constraints are lifted. For finite-rate channels, we employ gradient-based methods to compute the optimal controllers. Numerical experiments illustrate how tuning the privacy parameter shapes the trade-off between control performance and resilience against unauthorized inferences.

\end{abstract}

\section{INTRODUCTION}

In modern cyber-social-physical systems, information is a double-edged sword: it can be strategically released to influence decisions, yet excessive disclosure can compromise privacy and security. \emph{Bayesian persuasion} provides a formal game-theoretic framework to study how a knowledgeable sender (encoder) can design what information to reveal to a rational decision-maker (decoder) in order to influence the latter's action. Classic Bayesian persuasion models assume a single interested receiver, but many real-world scenarios involve additional {eavesdroppers} or adversarial observers who may intercept the communication. This raises fundamental questions about the trade-off between \textbf{strategic influence} and \textbf{privacy preservation}. For example, an IoT sensor in a smart grid may wish to signal the grid controller about a true system state \(x\) while hiding a private bias $\theta$ (e.g., calibration offset or proprietary parameter) from a malicious eavesdropper. Similarly, online platforms seek to disclose insights from user data to advertisers while protecting individual privacy. In all such cases, the encoder must balance providing enough information to sway the legitimate decoder's action with concealing sensitive information from unintended parties.

 We focus on models where an encoder has a random {private bias} parameter that should remain hidden from eavesdroppers even as the encoder influences the decoder's response to the true state (typically, the decoder acts based on \(x + \theta)\). We highlight theoretical results on how optimal signaling schemes trade off {influence} versus {privacy}.
 
When we introduce an eavesdropper into a persuasion model, the sender's problem becomes more complex. Now there are effectively multiple “receivers”: the intended decoder whose action the sender cares about, and an eavesdropper who should ideally learn as little as possible about the sender's private bias. The sender's utility typically includes a {privacy cost}. This scenario can be cast as a three-player game: the encoder (sender), the legitimate decoder, and the eavesdropper. A practical example is a remote state estimation system in a smart infrastructure: a sensor sends readings to a control center, while a malicious eavesdropper intercepts the communication to glean information about the system or the sensor's private parameters. In \cite{chen2024strategic}, this problem is formulated as a {tripartite game} with the sensor as a leader designing an optimal encoder, and the estimator and eavesdropper as followers who design their respective estimators. 

 Several theoretical models capture the trade-off between persuasion and privacy in such settings. In information economics, researchers have examined whether the concavification approach of \cite{kamenica2011bayesian, kamenica2019bayesian} extend to optimization over private signals in \cite{arieli2019private, babichenko2017algorithmic}.  A recent approach in \cite{pan2024differentially} studies differential privacy in the persuasion setting and quantifies how privacy requirements degrade the ability to persuade.
  In the context of cyber-physical system security, the sender-receiver interaction is often embedded in a control loop, and the eavesdropper may be an attacker who uses inferred information to harm the system, see e.g., \cite{farokhi2016privacy, akyol2015privacy}. Game-theoretic models treat the eavesdropper as an adversary, with the sensor deliberately adding artificial noise or encryption to its transmitted signals to confuse the eavesdropper.

In all such cases, the encoder must balance providing enough information to sway the legitimate decoder’s action with concealing sensitive information from unintended parties. In this paper, we particularly focus on the role of quantization in the privacy-persuasion tradeoff, building on the recent work on strategic quantization, see e.g., \cite{akyol2023isit, aybas2019persuasion, anand2025jsac}. 

Quantization, the process of discretizing a continuous signal into a finite set of values, is pervasive in modern communication systems. In a strategic communication context, quantization serves as both an engineering necessity and a deliberate design choice for privacy preservation. By restricting the resolution of the transmitted message, the sender can {pool} multiple states (or multiple potential values of the bias) into a single quantization bin, thereby confusing an eavesdropper about the exact underlying state.

However, this benefit comes at a cost. Fine quantization typically enables more accurate reconstruction of \(x \) by the legitimate receiver, but it also allows the eavesdropper to better infer the sender's private bias $\theta$. Thus, there exists an inherent trade-off: \textbf{coarse quantization} may enhance privacy at the expense of persuasion efficacy, whereas {fine quantization} improves the receiver's estimation at the risk of increased information leakage.

The set of admissible encoding mappings \( g(\cdot, \cdot) \) (or, equivalently, \( g(\cdot) \)) can be categorized into three distinct classes:
\begin{enumerate}[i)]
    \item \textbf{Non-revealing:} In this class, the sender \(\mathbf{S}\) transmits no information about the source, i.e.,
    \[
    g(X,\theta) = c,
    \]
    where \( c \) is a predetermined constant.
    
    \item \textbf{Fully revealing:} Here, the sender transmits the information exactly as requested by the receiver \(\mathbf{R}\), subject to the inherent rate constraints. Formally, this is represented as
    \[
    g(X,\theta) = Q(X),
    \]
    where \( Q(\cdot) \) denotes the channel-optimized non-strategic quantizer.
    
    \item \textbf{Partially revealing:} In this intermediate case, the sender conveys a message that is neither fully revealing nor completely non-revealing, yet still yields mutual benefit. That is,
    \[
    g(X,\theta) = Y,
    \]
    where \( Y \neq Q(X) \) and \( Y \neq c \).
\end{enumerate}

Our contributions in this paper are as follows:
\begin{enumerate}
    \item We derive the optimal mappings for privacy constrained strategic communication without rate constraints,
    \item We propose a design method for strategic quantization with  privacy constraints.
\end{enumerate}

Our analysis and numerical results have uncovered several important and rather surprising observations. The first one is that the presence of an eavesdropper helps the decoder in the strategic communication/quantization scenario. Our second observation pertains to quantization: the decoder might prefer quantization in the presence of a privacy constraints, although it prefers to use the communication channel fully when there is no privacy constraints. 
\begin{figure*}
    \centering
    \includegraphics[width=18cm]{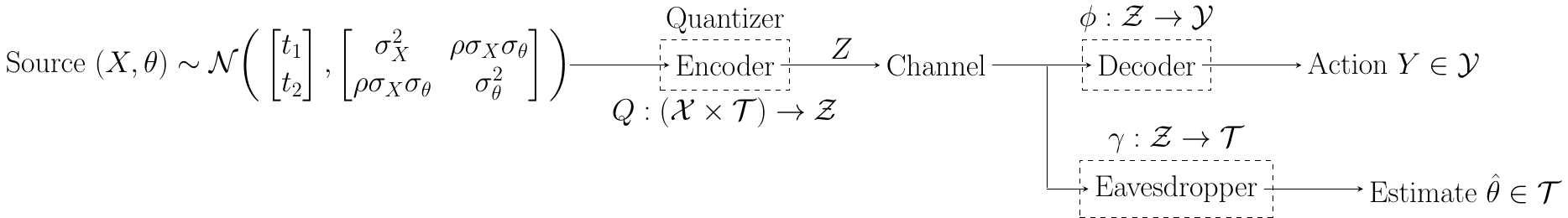}
    \caption{Communication diagram: $(X,\theta)$ over a noiseless channel with an eavesdropper}
    \label{fig:comm_diag}
\end{figure*}

This paper is organized as follows: In Section \rom{2} we present the problem formulation. In Section \rom{3}, we present the optimal mapping for the case with no rate constraints, and a gradient-descent based algorithm to compute the privacy-constrained strategic quantizer. We provide numerical results in \rom{4}, and conclude in \rom{5}.
\section{PRELIMINARIES}
\subsection{Notation}
In this paper, random variables are denoted using capital letters (say $X$), their sample values with respective lowercase letters ($x$), and their alphabet with respective calligraphic letters ($\mathcal{X}$). The set of real numbers is denoted by $\mathbb{R}$. The alphabet, $\mathcal{X}$, can be finite, infinite, or a continuum, like an interval $[a,b]\subset \mathbb R$. The  2-dimensional jointly Gaussian distribution with mean $\begin{bmatrix}
    t_1 & t_2
\end{bmatrix}'$ and respective variances $\sigma_1^2,\sigma_2^2$ with a correlation $\rho$ is denoted by  $\mathcal{N}\bigg (\begin{bmatrix}t_1\\ t_2    \end{bmatrix},\sigma_1^2\begin{bmatrix}
        1 &  \rho r \\  \rho r & r^2 
    \end{bmatrix}\bigg ), 0\leq \rho <1, r = \sigma_{2}/\sigma_1$, $t_1,t_2\in \mathbb{R}$, respectively.  The expectation operator is written as $\mathbb{E}\{\cdot\}$. The operator $|\cdot |$ denotes  the cardinality of the set.

\subsection{Problem Formulation}
Consider the following communication problem: an encoder observes realizations of the two sources $X\in \mathcal{X}\subseteq  [a_X,b_X],\theta \in \mathcal{T}\subseteq [a_{\theta},b_{\theta}]$, $a_X,b_X,a_{\theta},b_{\theta} \in \mathbb{R}$ with joint probability distribution $(X,\theta)\sim f(x,\theta)$, and maps $(X,\theta)$ to a message $Z\in \mathcal{Z}$, where $\mathcal{Z}$ is a set of  messages using a  mapping $Q:(\mathcal{X} \times \mathcal{T}) \rightarrow Z$. After receiving the message $Z$, the decoder applies a mapping $\phi:\mathcal{Z}\rightarrow \mathcal{Y}$ on the message $Z$ and takes an action $Y=\phi(Z) $. An eavesdropper observes the message $Y$ and estimates $\theta$ as $\hat{\theta} = \gamma(\theta)$.

The encoder, decoder, and eavesdropper minimize their respective objectives $$D_E=\mathbb{E}\{(X+\theta-Y)^2\} $$ such that $$\mathbb{E}\{(\theta - \hat{\theta})^2\}\geq C,$$ $C\in \mathbb{R}$,   $$D_D=\mathbb{E}\{(X-Y)^2\},$$ and $$D_{\theta} =\mathbb{E}\{(\theta - \hat{\theta})^2\}.$$ The encoder designs $Q$ \textit{ex-ante}, i.e., without the knowledge of the realization of $(X,\theta)$, using only the objectives $D_E$, $D_D$, $D_{\theta}$, and the statistics of the source $f(\cdot,\cdot)$. The objectives ($D_E$,  $D_D$, $D_{\theta}$), the shared prior ($f$), and the mapping ($Q$) are known to the encoder, decoder, and the eavesdropper. The problem is to design $Q$ for the equilibrium, i.e., the encoder minimizes its distortion if used with a corresponding decoder that minimizes its own distortion. This communication setting is given in Fig. \ref{fig:comm_diag}. Since the encoder chooses the mapping $Q$ first, followed by the decoder choosing the quantization representative levels ($\mathbf{y}$), we look for a Stackelberg equilibrium.

\section{MAIN RESULTS}
We assume a jointly Gaussian source $$(X,\theta)\sim \mathcal{N}\bigg(\begin{bmatrix}
    0 \\ 0
\end{bmatrix},\sigma_X^2 \begin{bmatrix}
    1 & \rho r \\ \rho r & r^2
\end{bmatrix}\bigg).$$ We consider two problem settings: one without quantization, and one with quantization. In the first case, we obtain closed form expressions for the mappings under some assumptions. In the second case, we provide an algorithm to compute the optimal quantizers.
\subsection{No quantization}

We make the following assumption on the mappings:
\begin{assumption}
    $Q(x,\theta) = x+\alpha \theta$, $\phi = \kappa y, \gamma= \nu y$, $\alpha,\kappa, \nu\in \mathbb{R}$.
\end{assumption}
We present the optimal mapping under these linearity assumptions below. The proof is
in Appendix \ref{appdx:mapping}.
\begin{theorem}
    The optimal mapping is given by 
    \begin{align}
    \alpha^* = \frac{ - (1+\lambda r^2)+ \sqrt{ (1+\lambda r^2)^2 - 4r(\rho + r)(\lambda \rho r -1) }  }{ 2r(\rho +r) }  \nonumber .
\end{align}
\label{thm:alpha}
\end{theorem}

\begin{remark}
    Although we obtain the solution for the general case of correlation $\rho$, in our numerical results we focus on $\rho = 0$ setting.
\end{remark}
As $\lambda \rightarrow \infty$, $\alpha = -\rho/r$, i.e., the encoder optimally estimates $\theta$ and removes this to make the term $(X+\alpha \theta)$ independent of $\theta$. For $\rho = 0$, as $\lambda \rightarrow \infty$, the encoder sends $X$  (fully-revealing).

The following theorem shows that the decoder may prefer the existence of an adversary. The proof is shown in Appendix \ref{appdx:decdistdecreasewithlambda}.
\begin{theorem}
    For a zero mean jointly Gaussian source with correlation $\rho = 0$, the decoder distortion decreases with respect to $\lambda$. 
    \label{thm:Ddec}
\end{theorem}

\begin{remark}
    While in the classical communication setup the existence of an adversary is against the objective of the decoder, in the strategic problem setting, the decoder benefits from the adversary. 
\end{remark}

For $\rho =0$, $\alpha\geq 0,$ $$ \frac{\partial \kappa}{\partial \alpha}\leq 0,$$ i.e., as $\alpha $ increases, $\kappa $ decreases. 
Since $$ \frac{\partial \alpha}{\partial \lambda}<0,$$ $\alpha $ decreases with $\lambda$, i.e., $\kappa$ increases with $\lambda$.  

For $\lambda \rightarrow \infty$, $\alpha = 0$, $\kappa =1$ (the decoder accepts the message as it is because the encoder is fully revealing). 

\subsection{With quantization}
Consider the problem setting where the cardinality of the message space is constrained, $|\mathcal{Z}|\leq M$. The encoder's mapping $Q$ is a non-injective mapping. The set $\mathcal{X}$ is divided into mutually exclusive and exhaustive sets parameterized by the realization of $\theta$ as $\mathcal{V}_{\theta,1},\mathcal{V}_{\theta,2},\ldots,\mathcal{V}_{\theta,M}$. The $m-$th quantization region is denoted by $\mathcal{V}_{:,m}=\{\mathcal{V}_{\theta,m},\forall \theta \in \mathcal{T}\}$. 
The encoder chooses the set of quantizers $Q=\{q_{\theta},\theta \in \mathcal{T}\}$  to minimize its distortion,
\begin{align}
    D_E& = \mathbb{E}\{(X+\theta - Y)^2\} - \lambda \mathbb{E}\{ (\theta - \hat{\theta})^2 \} \nonumber
    \\& =  \underset{m=1}{\overset{M}{\sum}} \underset{\theta\in \mathcal{T}}{\int} \underset{ x\in \mathcal{V}_{\theta,m} }{\int} (x+\theta - y_m)^2 \mathrm d  F(x,\theta) \nonumber
    \\& \quad -  \underset{m=1}{\overset{M}{\sum}} \underset{\theta\in \mathcal{T}}{\int} \underset{ x\in \mathcal{V}_{\theta,m} }{\int}\lambda (\theta - \hat{\theta}_m)^2\mathrm d  F(x,\theta), \label{eqn:DE}
\end{align}
where $\lambda$ is the Lagrangian parameter, and the optimal reconstruction points $y_m^*$ are determined by the decoder as a best response to $Q$ to minimize its distortion, 
\begin{align}
    y_m^*& = \underset{y \in \mathcal{Y}}{\argmin}\underset{m=1}{\overset{M}{\sum}} \mathbb{E}\{(X-y)^2|x\in \mathcal{V}_{:,m}\}\nonumber
    \\& = \frac{\underset{\theta\in \mathcal{T}}{\int} \underset{x \in \mathcal{V}_{\theta,m}}{\int} x \mathrm d  F(x,\theta) }{\underset{\theta\in \mathcal{T}}{\int} \underset{x \in \mathcal{V}_{\theta,m}}{\int} \mathrm d  F(x,\theta)  },\label{eqn:yval}
\end{align}
and the optimal estimates $\hat{\theta}_m$ are determined by the eavesdropper to minimize its distortion,
\begin{align}
    \hat{\theta}_m = \frac{  \underset{\theta\in \mathcal{T}}{\int} \underset{q_{\theta,m-1}}{\overset{q_{\theta,m}}{\int}} \theta \mathrm d  F(x,\theta) }{  \underset{\theta\in \mathcal{T}}{\int} \underset{q_{\theta,m-1}}{\overset{q_{\theta,m}}{\int}} \mathrm d  F(x,\theta) }. \label{eqn:thval}
\end{align}

\begin{figure}[h!tbp]
    \centering
    \includegraphics[width=8.5cm]{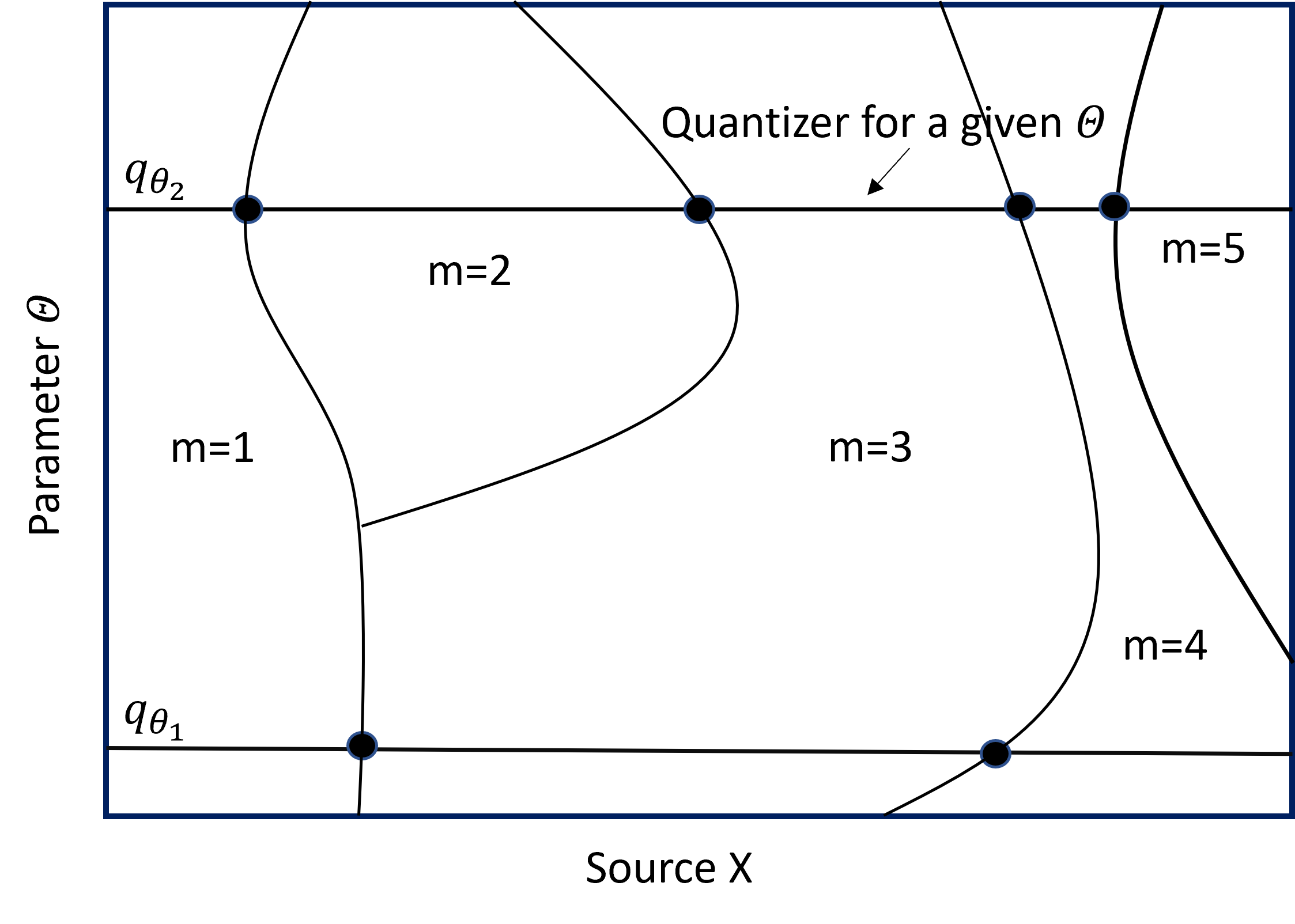}
    \caption{Quantization of $X$ parameterized by $\theta$ for $M=5$ illustrated.}
    \label{fig:quantM5}
\end{figure}

Note that implementing a quantizer $Q:(\mathcal{X},\mathcal {T})\rightarrow \mathcal{Z}$ can be simplified to computing a set of quantizers corresponding to each $\theta \in \mathcal{T}$ as in Fig. \ref{fig:quantM5} without loss of generality. If the quantizer does not include a region $m$ for some realization of $\theta$, the encoder never sends the message $m$ i.e., the encoder chooses a lower rate and is less revealing for that value of $\theta$. In Fig. \ref{fig:quantM5}, we see that the quantizer $q_{\theta_1}$ only includes $m=1,2,4$ regions, while the quantizer $q_{\theta_2}$ contains all five regions.

\begin{figure*}[htb]
\begin{minipage}[b]{0.32\linewidth}
  \centering
  \centerline{\includegraphics[width=6cm]{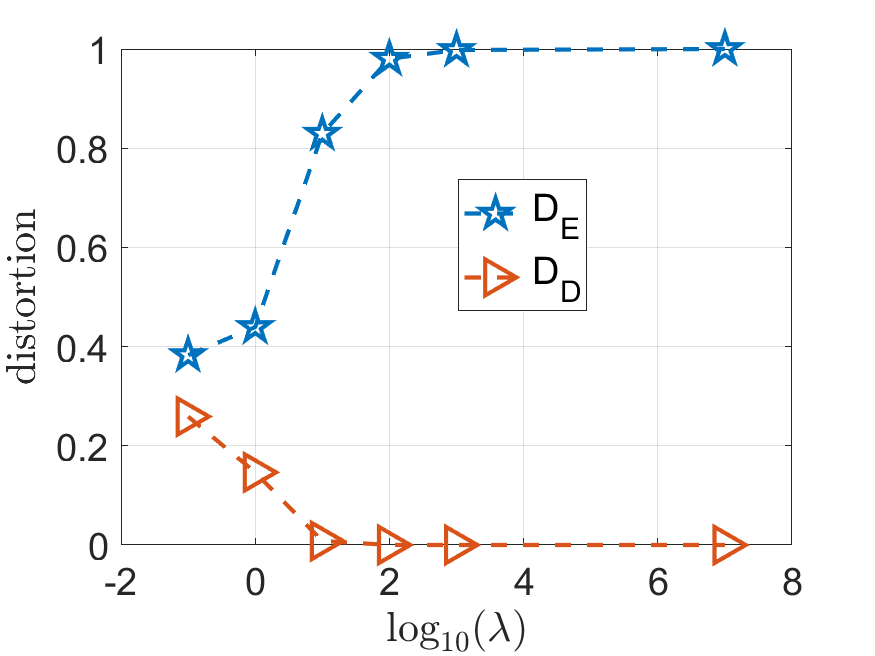}}
  \centerline{\scriptsize{(a) Without rate constraints}}
\end{minipage}
\hfill
\begin{minipage}[b]{.32\linewidth}
\label{fig:}
  \centering
  \centerline{\includegraphics[width=6cm]{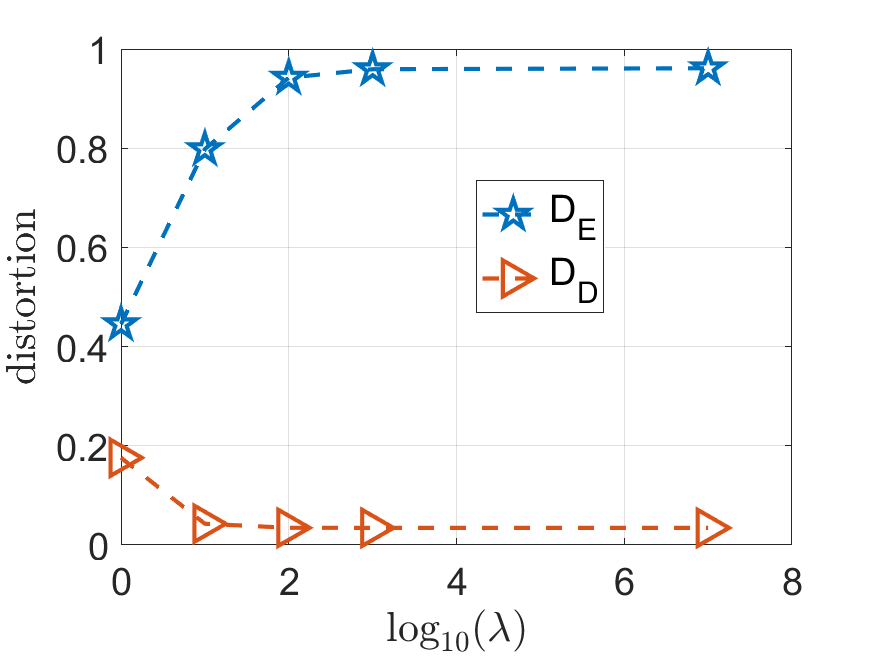}}
  \centerline{\scriptsize{(b) $M=8$}}
\end{minipage}
\hfill
\begin{minipage}[b]{.32\linewidth}
\label{fig:}
  \centering
  \centerline{\includegraphics[width=6cm]{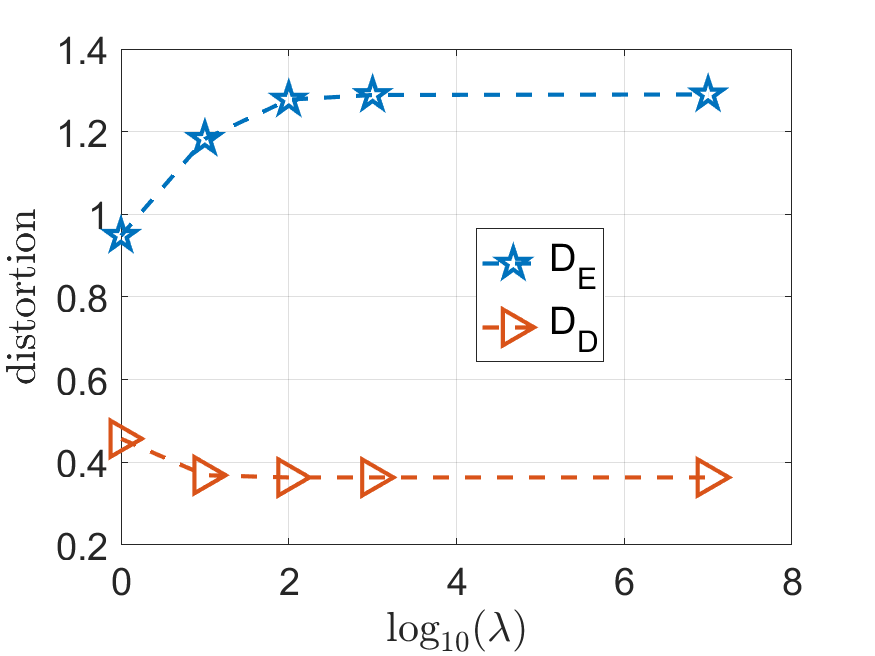}}
  \centerline{\scriptsize{(c) $M=2$}}
\end{minipage}
\caption{Distortions in quantizing a jointly Gaussian source $(X,\theta)\sim \mathcal{N}\bigg(\begin{bmatrix}
    0 \\ 0
\end{bmatrix}, \begin{bmatrix}
    1 & 0 \\ 0 & 1
\end{bmatrix}\bigg)$ with $D_E = \mathbb{E}\{(X+\theta - Y)^2,$ subject to $\mathbb{E}\{(\theta - \hat{\theta})^2\}\geq D_{\theta}$, $ D_D = \mathbb{E}\{(X-Y)^2\}$.
}
\label{fig:dist}
\end{figure*}

 \begin{figure*}[htb]
\begin{minipage}[b]{0.32\linewidth}
  \centering
  \centerline{\includegraphics[width=6cm]{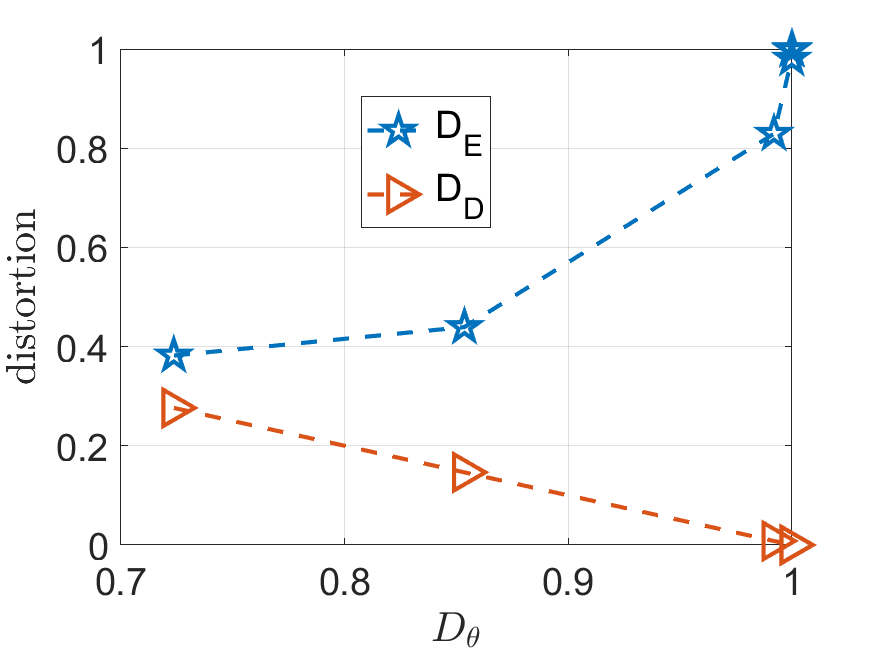}}
  \centerline{\scriptsize{(a) Without rate constraints}}
\end{minipage}
\hfill
\begin{minipage}[b]{.32\linewidth}
\label{fig:}
  \centering
  \centerline{\includegraphics[width=6cm]{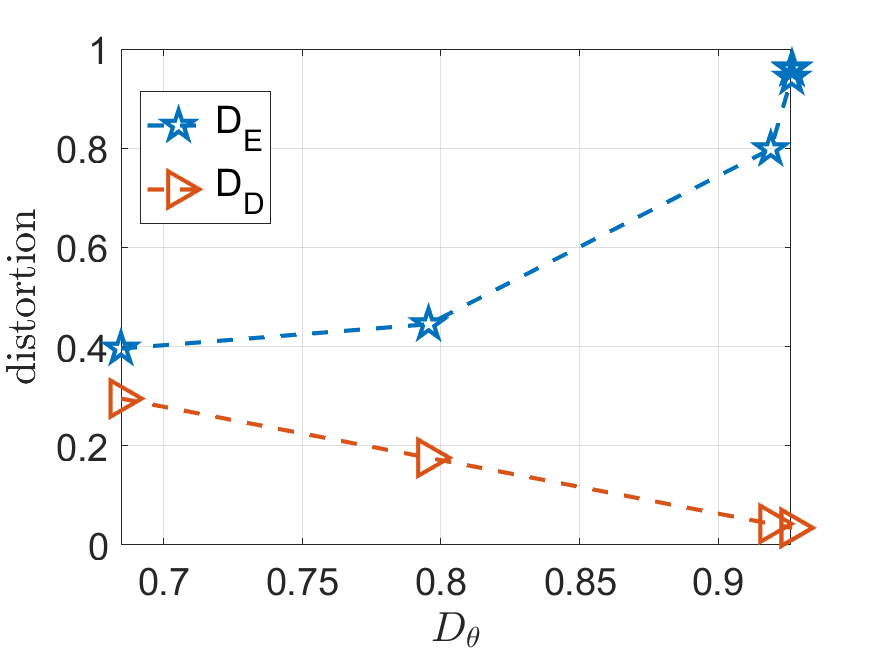}}
  \centerline{\scriptsize{(b) $M=8$}}
\end{minipage}
\hfill
\begin{minipage}[b]{.32\linewidth}
\label{fig:}
  \centering
  \centerline{\includegraphics[width=6cm]{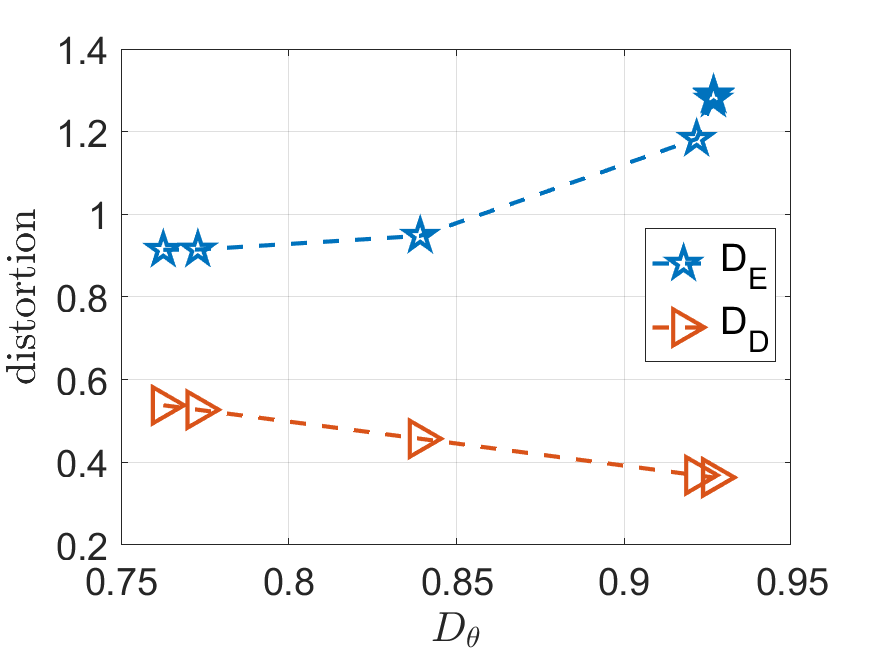}}
  \centerline{\scriptsize{(c) $M=2$}}
\end{minipage}
\caption{Distortions in quantizing a jointly Gaussian source $(X,\theta)\sim \mathcal{N}\bigg(\begin{bmatrix}
    0 \\ 0
\end{bmatrix}, \begin{bmatrix}
    1 & 0 \\ 0 & 1
\end{bmatrix}\bigg)$ with $D_E = \mathbb{E}\{(X+\theta - Y)^2,$ subject to $\mathbb{E}\{(\theta - \hat{\theta})^2\}\geq D_{\theta}$, $ D_D = \mathbb{E}\{(X-Y)^2\}$.
}
\label{fig:dist}
\end{figure*}

In \cite{akyol2023isit}, we proposed a gradient-descent based algorithm to solve the problem of quantization of a scalar source with misaligned encoder and decoder objectives communicating over a fixed rate noiseless channel. We extended this algorithm to a 2-dimensional source $(X,\theta)$ by a simple method of computing quantizers for each value of $\theta$ as $Q=\{q_{\theta},\theta \in \mathcal{T}\}, q_{\theta} $ in \cite{anand2024quadratic}. 

 Here, we extend this method to compute quantizer under privacy constraints. 
The gradient descent optimization is performed with the objective as the encoder distortion optimized over the encoder's choice of quantizer decision levels $Q=\{q_{\theta},\theta \in \mathcal{T}\}$. Although the encoder distortion depends on decoder reconstruction levels $\mathbf{y}$, since $\mathbf{y}$ is a function of $Q$, the optimization can be implemented as a function of solely $Q$.

\begin{algorithm}[h!tbp]
\caption{Proposed strategic quantizer design} 
 Parameters: $\epsilon,\eta$\\
 Input: $f(\cdot,\cdot),\mathcal{X},\mathcal{T},M$\\
 Output: $\{q_{\theta}^*\}$, $\{y_m^*\}$, $\{\hat{\theta}_m^*\}$, $D_E$, $D_D$, $D_{\theta}$\\
 Initialization: assign a set of monotone $\{q_{\theta,0}\}$ randomly, compute associated encoder distortion $D_E(0)$, set iteration index $i=1$;\\
\While{ $\Delta D >\epsilon $ or until a set amount of iterations} {
    compute the gradients $\{\partial D_E /\partial x_{\theta,:}\}_i$, \\
    compute the updated quantizer  $q_{\theta,i+1} \triangleq q_{\theta,i} - \eta \{\partial D_E /\partial x_{\theta,:}\}_i$ for $\theta \in \mathcal{T}$, \\
    compute actions  $\mathbf{y}(\{q_{\theta,i+1}\})$ via (\ref{eqn:yval}),\\
    compute estimates  $\mathbf{\hat{\theta}}(\{q_{\theta,i+1}\})$ via (\ref{eqn:thval}),\\
    compute encoder distortion $D_E (i+1)$ associated with quantizer values $q_{\theta,i+1}$, actions $\mathbf{y}(\{q_{\theta,i+1}\})$, and estimates $\mathbf{\hat{\theta}}(\{q_{\theta,i+1}\})$ via (\ref{eqn:DE}),\\
    compute  $\Delta D =D_E (i)-D_E (i+1)$. 
   }  
\Return quantizer $\{q_{\theta}^*\}=\{q_{\theta,i+1}\}$, actions $\{y_m^*\}=\mathbf{y}(\{q_{\theta}^*\})$, estimates $\{\hat{\theta}^*_m\} = \mathbf{\hat{\theta}}(\{q_{\theta}^*\})$, encoder, decoder, and eavesdropper distortions $D_E$, $D_D$, and $D_{\theta}$ computed for optimal quantizer and decoder actions $\{q_{\theta}^*\},\mathbf{y}(\{q_{\theta}^*\}),\mathbf{\hat{\theta}}(\{q_{\theta}^*\})$ via (\ref{eqn:DE}).  
\end{algorithm}
\section{NUMERICAL RESULTS}
We present results for the following three settings 
\begin{enumerate}
    \item No rate constraint ($M\rightarrow \infty$)
    \item $M=8$
    \item $M=2$
\end{enumerate}
for a jointly Gaussian source $$(X,\theta)\sim \mathcal{N}\bigg(\begin{bmatrix}
    0 \\ 0
\end{bmatrix}, \begin{bmatrix}
    1 & 0 \\ 0 & 1
\end{bmatrix}\bigg)$$ with $$D_E = \mathbb{E}\{(X+\theta - Y)^2 $$ subject to $\mathbb{E}\{(\theta - \hat{\theta})^2\}\geq D_{\theta}$, $$D_D = \mathbb{E}\{(X-Y)^2\}.$$

In Figure \ref{fig:dist}, we present the encoder, decoder, and eavesdropper distortions. As $\lambda$ increases, the encoder’s distortion grows, whereas the decoder’s distortion decreases. In other words, enhancing privacy requirements diminishes the encoder’s ability to persuade, as we intuitively expect. 

\begin{figure*}
\begin{minipage}[b]{0.32\linewidth}
  \centering
  \centerline{\includegraphics[width=6cm]{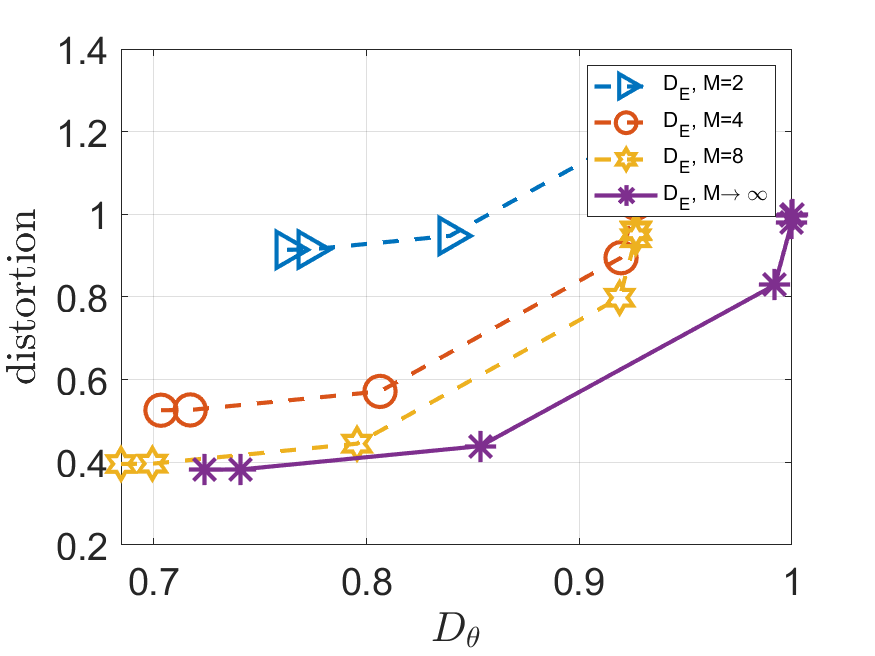}}
  \centerline{\scriptsize{(a) $D_E$}}
\end{minipage}
\hfill
\begin{minipage}[b]{0.32\linewidth}
  \centering
  \centerline{\includegraphics[width=6cm]{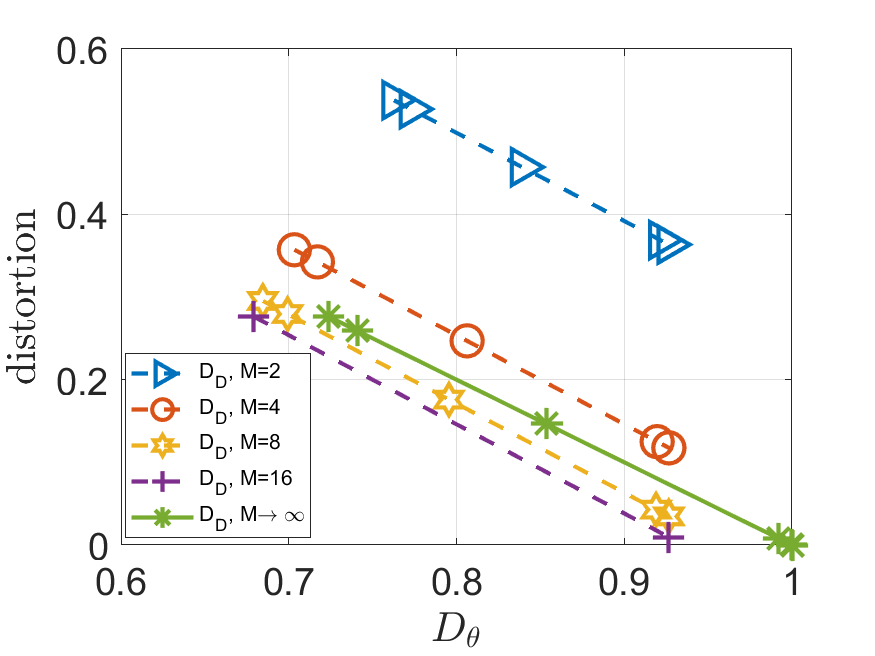}}
  \centerline{\scriptsize{(b) $D_D$}}
\end{minipage}
\hfill
\begin{minipage}[b]{0.32\linewidth}
  \centering
  \centerline{\includegraphics[width=6cm]{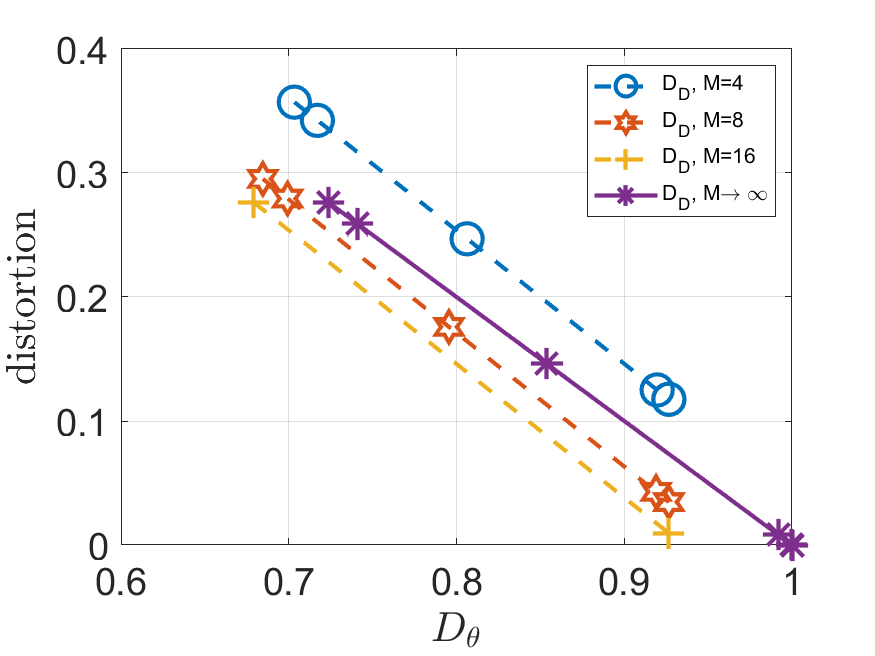}}
  \centerline{\scriptsize{(c) $D_D$ zoomed in}}
\end{minipage}
\caption{Distortions in quantizing a jointly Gaussian source $(X,\theta)\sim \mathcal{N}\bigg(\begin{bmatrix}
    0 \\ 0
\end{bmatrix}, \begin{bmatrix}
    1 & 0 \\ 0 & 1
\end{bmatrix}\bigg)$ with $D_E = \mathbb{E}\{(X+\theta - Y)^2,$ subject to $\mathbb{E}\{(\theta - \hat{\theta})^2\}\geq D_{\theta}$, $ D_D = \mathbb{E}\{(X-Y)^2\}$ for different $M$ values.
}
\label{fig:distortions1}
\end{figure*}

Figure \ref{fig:distortions1} illustrates that, although the encoder generally prefers higher rates, the decoder may unexpectedly benefit from a certain degree of quantization rather than having no quantization at all. This preference arises because quantization introduces a level of privacy, partially satisfying the encoder’s privacy objective. Hence, the encoder is  inclined to reveal more information about 
the state $X$ which better serves the decoder’s interests. Numerical results indicate that there may be an optimal rate from the decoder’s standpoint, and we leave a more extensive analysis of this rather surprising phenomenon for future work.

\begin{figure}
    \centering
    \includegraphics[width=0.8\linewidth]{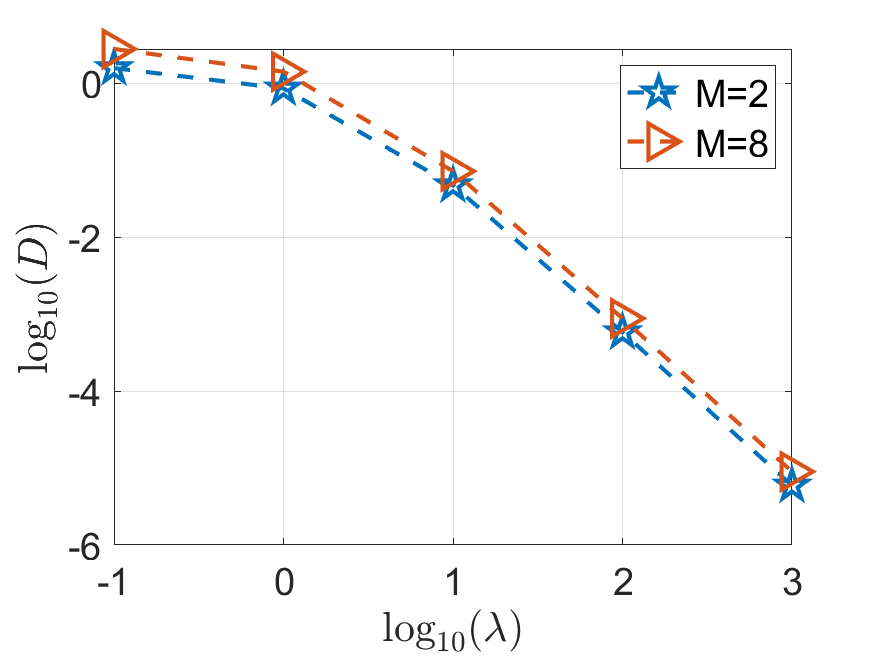}
    \caption{Similarity measure $D$ of the optimal strategic quantizer under privacy constraints.}
    \label{fig:KL}
\end{figure}

We define a measure of similarity between quantizers corresponding to each $\theta \in \mathcal{T}$ using KL divergence as follows:
\begin{align}
    D_{KL}(\theta_1, \theta_2) = \underset{m=1}{\overset{M}{\sum}} \underset{q_{\theta_1,m-1}}{\overset{q_{\theta_1,m}}{\int}} \mathrm d F(x) \log \frac{ \underset{q_{\theta_1,m-1}}{\overset{q_{\theta_1,m}}{\int}} \mathrm d F(x) }{\underset{q_{\theta_2,m-1}}{\overset{q_{\theta_2,m}}{\int}} \mathrm d F(x) }\nonumber . 
\end{align}
For a quantizer $\{q_{\theta,m},\theta \in \mathcal{T},m\in [1:M]\}$, the similarity of the quantizer is given by
\begin{align}
    D = \underset{\theta_1,\theta_2\in \mathcal{T}}{\max} D_{KL}(\theta_1,\theta_2). \nonumber
\end{align}
We plot $D$ in Fig. \ref{fig:KL} for $M=2,8$ quantizer for different $\lambda $ values, and we observe that $D$ decreases with $\lambda$, i.e.,  the quantizers are increasingly similar as $\lambda$ increases or as the encoder's constraint on privacy becomes more stringent.

Finally, as \(\lambda \to \infty\), the decoder’s distortion converges to the fully revealing scenario in all three settings: \(0\) for no rate constraint, \(0.0345\) for \(M = 8\), and \(0.3634\) for \(M = 2\). This occurs because, under extremely strict privacy requirements, the encoder focuses exclusively on meeting the privacy constraint and no longer prioritizes minimizing
\[
\mathbb{E}\bigl\{ (X + \theta - Y)^2 \bigr\}.
\]

The decoder distortion as $\lambda \rightarrow\infty$ approaches the fully-revealing decoder distortion in the three cases (0 for no rate constraint, 0.0345 for $M=8$, 0.3634 for $M=2$). This is because the encoder is concerned solely about the privacy constraint and does not optimize $\mathbb{E}\{(X+\theta-Y)^2\}$.

As $\lambda \rightarrow \infty$, 
$\alpha = -\rho/r = 0$, 
\begin{align}
    D_E & = \mathbb{E}\{(X+\theta-Y)^2\}
    \\& = \mathbb{E}\{(X-Y)^2\} + \mathbb{E}\{\theta^2\} + 2\mathbb{E}\{\theta(X-Y)\}
\end{align}
which evaluates to the following for $\rho = 0$,
\begin{align}
    D_E& = 
    \mathbb{E}\{(X-Y)^2\} + \mathbb{E}\{\theta^2\},
\end{align}
and we observe the encoder distortion for $\lambda\rightarrow \infty$ as $D_D+ \mathbb{E}\{\theta^2\}$ where $\mathbb{E}\{\theta^2\}$ is computed numerically  
in Fig. \ref{fig:dist}.

\begin{figure*}[htb]
\begin{minipage}[b]{0.32\linewidth}
  \centering
  \centerline{\includegraphics[width=6cm]{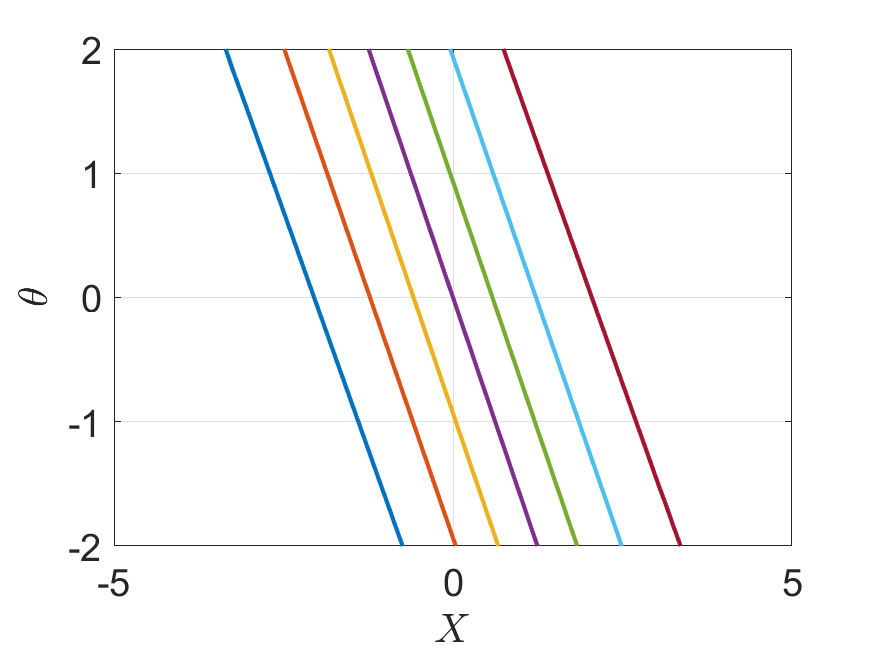}}
  \centerline{\scriptsize{(a) $\lambda = 0$}}
\end{minipage}
\hfill
\begin{minipage}[b]{.32\linewidth}
\label{fig:}
  \centering
  \centerline{\includegraphics[width=6cm]{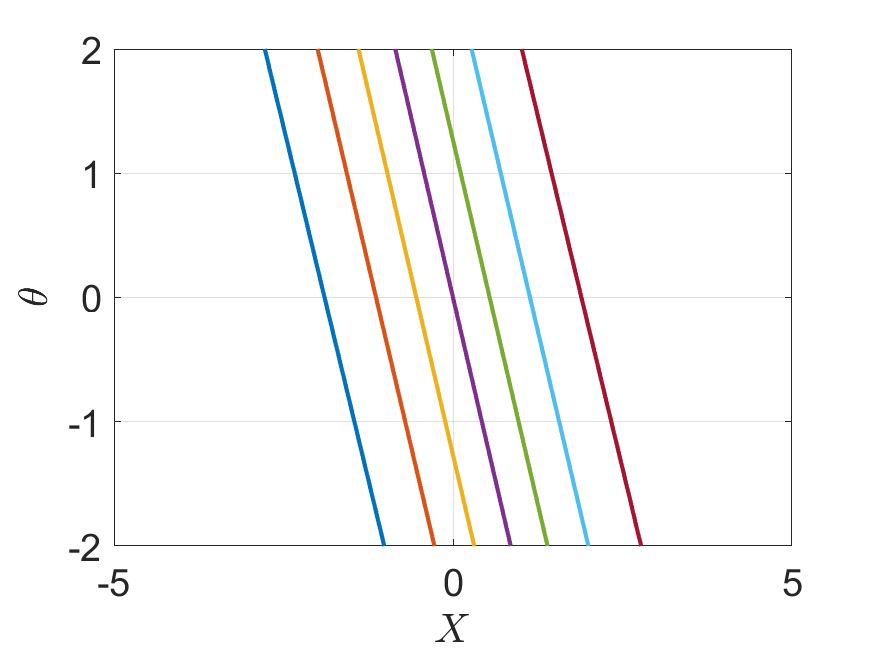}}
  \centerline{\scriptsize{(b) $\lambda = 1$}}
\end{minipage}
\hfill
\begin{minipage}[b]{.32\linewidth}
\label{fig:}
  \centering
  \centerline{\includegraphics[width=6cm]{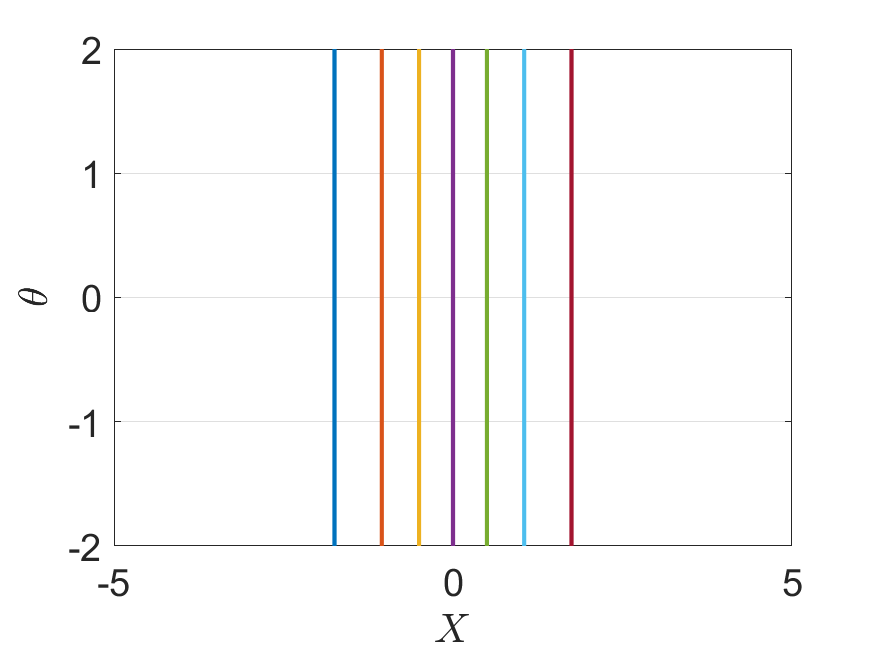}}
  \centerline{\scriptsize{(c) $\lambda = 10^7$}}
\end{minipage}
\caption{Quantizers for a jointly Gaussian source $(X,\theta)\sim \mathcal{N}\bigg(\begin{bmatrix}
    0 \\ 0
\end{bmatrix}, \begin{bmatrix}
    1 & 0 \\ 0 & 1
\end{bmatrix}\bigg)$ with $D_E = \mathbb{E}\{(X+\theta - Y)^2,$ subject to $\mathbb{E}\{(\theta - \hat{\theta})^2\}\geq D_{\theta}$, $ D_D = \mathbb{E}\{(X-Y)^2\}$.
}
\label{fig:quantizers}
\end{figure*}

As $\lambda \rightarrow \infty$, the encoder becomes fully revealing, as we observe in the quantizers in Fig. \ref{fig:quantizers}. We note that from numerical results the optimal quantizers appear to gradually shift to the fully revealing one, as opposed to being fully revealing after a certain threshold of $\lambda$.


\section{CONCLUSIONS}

In this paper, we analyzed the problem of privacy-constrained strategic communication of a 2-dimensional source $(X,\theta)$ with encoder objective to minimize $\mathbb{E}\{(X+\theta-Y)^2\}$ such that $\mathbb{E}\{(\theta - \hat{\theta})^2\}\geq C$, where $C$ is a constant, decoder objective to minimize $D_D = \mathbb{E}\{(X-Y)^2\}$, and the eavesdropper objective to minimize $D_{\theta} = \mathbb{E}\{(\theta - \hat{\theta})^2\}$. We extended a prior result on strategic communication to the privacy constrained case. We incorporated rate constraints, and presented an algorithm for the design of the strategic quantizer under the privacy constraint. The numerical results obtained suggest several intriguing research problems which we leave as a part of our future work.

\appendices
\section{Proof of Theorem \ref{thm:alpha}}
\label{appdx:mapping}
We have two scalar random variables $X$ and $\theta$, jointly Gaussian with zero means and covariance
\begin{align}
    (X,\theta)\sim \mathcal{N}\bigg(\begin{bmatrix}
    0 \\ 0
\end{bmatrix},\begin{bmatrix}
    \sigma_X^2  & \rho \sigma_X\sigma_{\theta} \\ \rho \sigma_X\sigma_{\theta} & \sigma_{\theta}^2 
\end{bmatrix}\bigg)\nonumber ,
\end{align}
where $\rho\in[-1,1]$. We form the scalar
$$y = x + \alpha \theta,$$
and use the linear (MMSE) estimates
$$\hat{X}(Y)
=
\frac{\mathbb{E}\{XY\}}{\mathbb{E}\{Y^2\}}Y,
\quad
\hat{\theta}(Y)
=
\frac{\mathbb{E}\{\theta Y\}}{\mathbb{E}\{Y^2\}}Y.$$
We want to choose $\alpha$ to minimize
$$J(\alpha)
=
\mathbb{E}\{(X +\theta-\hat{X}(Y))^{2}\}
-
\lambda \mathbb{E}\{(\theta -\hat{\theta}(Y))^{2}\}
\quad
(\lambda>0).$$

We define:
$$v(\alpha)
=
\mathbb{E}\{Y^2\}
=\mathbb{E}\{(X+\alpha \theta)^2\}
=
\sigma_X^2
+
2\alpha \rho \sigma_X \sigma_{\theta}
+
\alpha^2 \sigma_{\theta}^2.$$
$$c_x(\alpha)
=\mathbb{E}\{XY\}
=\mathbb{E}\{X(X+\alpha \theta)\}=
\sigma_X^2
+
\alpha \rho \sigma_X \sigma_{\theta},$$
$$c_s(\alpha)
=\mathbb{E}\{\theta Y\}
=\mathbb{E}\{\theta(X+\alpha \theta)\}
=
\rho\sigma_X \sigma_{\theta}
+
\alpha \sigma_{\theta}^2,$$
\begin{align}
    c_{x+\theta}(\alpha)
&=\mathbb{E}\{(X+\theta)Y\}
=\mathbb{E}\{(X+\theta)(X+\alpha \theta)\}\nonumber
\\&=
\sigma_X^2 + \rho \sigma_X\sigma_{\theta}
+
\alpha (\rho \sigma_X \sigma_{\theta} + \sigma_{\theta}^2).\nonumber
\end{align}

We write
$$
J(\alpha)
 = 
\mathbb{E} \{(X - \hat{X}(Y) +\theta)^{2} \}
 - 
\lambda 
\mathbb{E} \{(\theta - \hat{\theta}(Y))^{2} \}.
$$
After some straightforward algebra steps, we obtain:

$$
J(\alpha)
=
\underbrace{\sigma_X^2  + (1-\lambda) \sigma_{\theta}^2  + 2 \rho \sigma_X \sigma_{\theta}}_{\text{constant}}
 + 
\frac{P(\alpha)}{v(\alpha)}.
$$
where
$$
P(\alpha)
 = 
(c_x(\alpha))^{2}
 - 
2 c_x(\alpha) c_{x+\theta}(\alpha)
 + 
\lambda (c_s(\alpha))^{2}.
$$
Minimizing $J(\alpha)$ is therefore equivalent to minimizing
$\displaystyle \frac{P(\alpha)}{v(\alpha)}.$

We then expand the terms in $P(\alpha)$ and after some straightforward algebra (omitted here, presented in \cite{anand2025arxivprivacy}), we have a quadratic equation in terms of $\alpha$:

$$
r(\rho + r) \alpha^{2}
 + 
(1 + \lambda r^{2}) \alpha
 + 
(\lambda \rho r  - 1)
 = 
0, 
$$
where
$$
r  = \frac{\sigma_{\theta}}{\sigma_X}.
$$

Solving for  $\alpha$:
$$
\alpha^{\ast}
 = 
\frac{
-(1 + \lambda r^{2})
 \pm 
\sqrt{(1 + \lambda r^{2})^{2}
       - 
      4 r(\rho + r) (\lambda \rho r -1)}
}{
2 r (\rho + r)}.
$$

The second derivative,
\begin{align}
    \frac{\partial^2 }{\partial \alpha^2}\bigg[ \frac{P(\alpha)}{v(\alpha)}\bigg] & = 2r(\rho+r )\alpha + (1+\lambda r^2) \nonumber
    \end{align}

For $r>0,\, \rho\in[-1,1],\,\lambda >0$,
this expression evaluates to a negative value for 
$$\alpha  <-\frac{(1+\lambda r^2)}{2r(\rho+r )}.$$

That is,
$$
\alpha^{\ast}
 = 
\frac{
-(1 + \lambda r^{2})
+ 
\sqrt{(1 + \lambda r^{2})^{2}
       - 
      4 r(\rho + r) (\lambda \rho r -1)}
}{
2 r (\rho + r)}.
$$

\section{Proof of Theorem \ref{thm:Ddec}}
\label{appdx:decdistdecreasewithlambda}
We assume zero mean, $\rho = 0$.
\begin{align}
    D_D&= \mathbb{E}\{(X-\kappa(X+\alpha\theta))^2\}\nonumber
    \\& = (1-\kappa)^2 \mathbb{E}\{X^2\} + \kappa^2 \alpha^2 \mathbb{E}\{\theta^2\}\nonumber .
\end{align}
The term $\partial D_D/\partial \lambda $,
\begin{align}
    \frac{\partial D_D}{\partial \lambda } = \frac{\partial D_D}{\partial \alpha} \frac{\partial \alpha}{\partial \lambda}\nonumber .
\end{align}
We  define the term $f(\alpha)$ as follows for convenience:
\begin{align}
    f(\alpha) = \frac{ \sigma_X^2 }{ \sigma_X^2 + \alpha^2 \sigma_S^2 }.\nonumber
\end{align}
Then, the decoder distortion,
\begin{align}
    D_D & = (1-f(\alpha))^2 \mathbb{E}\{X^2\} + f(\alpha)^2 \alpha^2 \mathbb{E}\{\theta^2\}.\nonumber
\end{align}
The term $f'(\alpha)$ is given by,
\begin{align}
    \frac{\partial f(\alpha)}{\partial \alpha} & = -\sigma_X^2 (\sigma_X^2+\alpha^2 \sigma_S^2)^{-2} 2 \alpha \sigma_S^2 . \nonumber
\end{align}
The term $\partial D_D/\partial \alpha$,
\begin{align}
    \frac{\partial D_D}{\partial \alpha} & = -2(1-f(\alpha)) f'(\alpha) \mathbb{E}\{X^2\} \nonumber
    \\& \quad + 2 f(\alpha) f'(\alpha) \alpha^2 \mathbb{E}\{\theta^2\} +2f(\alpha)^2 \alpha \mathbb{E}\{\theta^2\}  \nonumber
\end{align}

We define $A,B$ as follows for convenience of notation:
\begin{align}
    A = 1+\lambda r^2\nonumber
\end{align}
\begin{align}
    B = (1+\lambda r^2)^2 - 4 r(\rho+r)(\lambda \rho r -1).\nonumber
\end{align}
Then $\alpha^*$ can be written as,
\begin{align}
    \alpha^* = \frac{-A+ \sqrt{B}  }{ 2 r (\rho+r)}.\nonumber
\end{align}
The term $\partial \alpha /\partial \lambda$,
\begin{align}
    \frac{\partial \alpha }{\partial \lambda} = \frac{ -\frac{\partial A}{\partial \lambda} + \frac{1}{2} B^{-1/2} \frac{\partial B}{\partial \lambda } }{ 2 r (\rho +r ) },\nonumber
\end{align}
\begin{align}
    \frac{\partial A}{\partial \lambda} = r^2\nonumber
\end{align}
\begin{align}
     \frac{\partial B}{\partial \lambda} =2(1+\lambda r^2) r^2 -4r(\rho +r)\rho r \nonumber
\end{align}
\begin{align}
      \frac{\partial \alpha }{\partial \lambda} = \frac{ -r^2 + \frac{1}{2} B^{-1/2} (2(1+\lambda r^2) r^2 -4r(\rho +r)\rho r ) }{ 2 r (\rho +r ) }\nonumber .
\end{align}

For $\rho = 0$:
\begin{align}
    \frac{ \partial \alpha }{ \partial \lambda } & = \frac{ -r^2 + \frac{(1+\lambda r^2)r^2}{  \sqrt{(1+\lambda r^2 )^2 + 4r^2}  } }{ 2 r^2} \nonumber
    \\& = -\frac{1}{2} + \frac{1}{2} \frac{ (1+\lambda r^2) }{ \sqrt{(1+\lambda r^2)^2 + 4 r^2} }\nonumber
    \\& = -\frac{1}{2} + \frac{1}{2} k\nonumber
\end{align}
where $k<1$, i.e., $$ \frac{ \partial \alpha }{ \partial \lambda }<0.$$

Let $T$ be the sum of the first two terms of $\partial D_D /\partial \alpha$,
\begin{align}
   T& =  -2(1-f(\alpha)) f'(\alpha) \mathbb{E}\{X^2\} + 2 f(\alpha) f'(\alpha) \alpha^2 \mathbb{E}\{\theta^2\} \nonumber\\ & = 2f'(\alpha) \bigg( -(1-f(\alpha)) \mathbb{E}\{X^2\} + f(\alpha) \alpha^2 \mathbb{E}\{\theta^2\} \bigg)\nonumber
    \\& = \frac{2f'(\alpha) }{ \sigma_X^2 + \alpha^2 \sigma_{\theta}^2 } \bigg( - \alpha^2 \sigma_{\theta}^2 \mathbb{E}\{X^2\} + \sigma_X^2 \alpha^2 \mathbb{E}\{\theta^2\} \bigg)\nonumber
\end{align}
Since we assumed zero mean,
\begin{align}
    T = 0.\nonumber
\end{align}
Then $\partial D_D/\partial \alpha$ evaluates to
\begin{align}
    \frac{\partial D_D}{\partial \alpha} & = 2f(\alpha)^2 \alpha \mathbb{E}\{\theta^2\}  .\nonumber
\end{align}
For $\rho = 0$, 
\begin{align}
    \alpha & = \frac{ -(1+\lambda r^2) + \sqrt{(1+\lambda r^2)^2 - 4 r(\rho+r)(\lambda \rho r -1)} }{ 2 r(\rho +r)}\nonumber
    \\& =  \frac{ -(1+\lambda r^2) + \sqrt{(1+\lambda r^2)^2 + 4 r^2} }{ 2 r^2}\nonumber
    \\& \geq 0\nonumber .
\end{align}
Then $\frac{\partial D_D}{\partial \alpha}\geq 0$, that is $\frac{\partial D_D }{\partial \lambda } \leq 0$, $D_D$ decreases as $\lambda$ increases.

\bibliographystyle{IEEEtran}
\bibliography{ref12_arxiv}
\end{document}